\documentclass[twocolumn, trackchanges]{aastex631}

\usepackage{amsmath}

\newcommand{\Deltatheta}{\overline{\Delta\theta}}

\newcommand{\vthetaPlain}{v_{d\theta}}
\newcommand{\vtheta}{\overline{v_{d\theta}}}
\newcommand{\vthetaScat}{v_{d\theta,scat}}
\newcommand{\vthetaTurb}{v_{d\theta,turb}}
\newcommand{\vthetaTheor}{v_{d\theta,theor}}

\definecolor{CommentRed}{RGB}{228, 0, 0}

\begin{document}

\title{Interplay of large-scale drift and turbulence in the heliospheric propagation of solar energetic particles}

\author[0000-0002-7719-7783]{T. Laitinen} 
\affiliation{Jeremiah Horrocks Institute, University of Central Lancashire, UK}
\author[0000-0002-7837-5780]{S. Dalla} 
\affiliation{Jeremiah Horrocks Institute, University of Central Lancashire, UK}


\begin{abstract}
The gradient and curvature of the Parker spiral interplanetary magnetic field give rise to curvature and gradient guiding centre drifts on cosmic rays. The plasma turbulence present in the interplanetary space is thought to suppress the drifts, however the extent to which they are reduced is not clear.
    We investigate the reduction of the drifts using a new analytic model of heliospheric turbulence where the dominant 2D component has both the wave vector and the magnetic field vector normal to the Parker spiral, thus fulfilling the main criterion of 2D turbulence.
    We use full-orbit test particle simulations of energetic protons in the modelled interplanetary turbulence, and analyse the mean drift velocity of the particles in heliolatitude. We release energetic proton populations of 10, 100 and 1000~MeV close to Sun and introduce a new method to assess their drift. We compare the drift in the turbulent heliosphere to drift in a configuration without turbulence, and to theoretical estimates of drift reduction.
    We find that  drifts are reduced by a factor 0.2-0.9 of that expected for the heliospheric configuration without turbulence. This corresponds to a  much less efficient suppression than what is predicted by theoretical estimates, particularly at low proton energies.
    We conclude that guiding centre drifts are a significant factor for the evolution of cosmic ray intensities in the heliosphere including the propagation of solar energetic particles in the inner heliosphere.
\end{abstract}

\keywords{Sun: particle emission - Sun: heliosphere - magnetic fields - turbulence - methods: numerical}

\section{Introduction} \label{sec:intro}

The heliosphere is traversed by different populations of energetic charged particles, the generally termed cosmic rays (CRs), with sources varying from the Sun and interplanetary space to outside the heliosphere in galactic and extragalactic sources. The propagation of these particles is guided by the interplanetary magnetic field (IMF), which has the macroscopic shape of an Archimedean spiral, the Parker spiral, due to the magnetic field, originating from the rotating Sun, being frozen in the solar wind plasma \citep{Parker1958_DynamicsInterplanetaryGas}.

In the simplest approximation, the CRs propagate parallel to the Parker spiral magnetic field. However, the magnetic field is curved, and its magnitude depends on the heliocentric distance, thus the CRs are subject to gradient and curvature guiding centre drifts \citep[e.g.][]{Burns1968_DynamicsChargedParticle}. The large-scale drifts influence the modulation of  the intensities of galactic cosmic rays (GCRs) that propagate through the outer heliosphere to be observed at Earth \citep[e.g.][]{Jokipii1977_EffectsParticleDrift}. They also cause solar energetic particles (SEPs) to drift in latitude and longitude, and lose energy \citep[e.g.][]{Dalla2013_SolarEnergeticParticle,Marsh2013_Drift-inducedPerpendicularTransport, Dalla2015_Drift-inducedDecelerationSolar}.

The large-scale IMF is superposed by a fluctuating component, which is due to the solar wind turbulence. This turbulent component causes field line random-walk resulting in CRs spreading stochastically across the large-scale average field. Also the velocity vector of the CRs is affected by the turbulence, which results in scattering of the CRs along the random-walking field lines \citep{Parker1965_PassageEnergeticCharged,Jokipii1966_Cosmic-RayPropagationI}. 

The interplay between the effects of turbulence and the drift motion due to large scale gradients and curvature has gained significant attention, however, details of this interaction are not clear. For GCRs, within modulation models large-scale drifts need to be suppressed in order for the models to be able to reproduce the observations \citep[e.g.][]{Potgieter1989_Interplanetarycosmicray}. Several theoretical works have suggested that turbulence reduces large-scale drifts \citep[e.g.][]{Gleeson1969_equationsdescribingcosmic-ray, Forman1974_Cosmic-RayStreamingPerpendicular,Bieber1997_PerpendicularDiffusionDrift, Giacalone1999_ParticleDriftsFluctuating, Engelbrecht2017_TowardGreaterUnderstanding, Vandenberg_2021_TurbulentReductionDrifts}, and reduction of drifts has found some support from full-orbit test particle simulations of charged particles in synthetic turbulent magnetic fields \citep[e.g.][]{Giacalone1999_ParticleDriftsFluctuating, Candia2004_Diffusiondriftcosmic, Minnie2007_SuppressionParticleDrifts, Tautz2012_DriftCoefficientsCharged}. However, these theoretical works and simulation studies have used either a constant or gradient-only background magnetic field configuration: thus, possible suppression of CR drifts in a realistic heliospheric context has not been probed.

In this work, we investigate the effect of turbulence on guiding centre drifts of solar energetic particles (SEPs) by means of 3D test particle simulations including turbulence superposed on a Parker spiral IMF. We make use of our newly-developed analytical model of composite plasma turbulence in the Parker spiral heliospheric configuration \citep{Laitinen2023_AnalyticalModelTurbulence}. We compare the latitudinal drift of SEPs in the turbulent heliosphere to that in IMF without turbulence \citep{Marsh2013_Drift-inducedPerpendicularTransport, Dalla2013_SolarEnergeticParticle}. We compare simulation results with the predictions of drift reduction models \citep{Bieber1997_PerpendicularDiffusionDrift, Engelbrecht2017_TowardGreaterUnderstanding}, and discuss the implications of our results on the theoretical models of drift reduction.

\section{Methods} \label{sec:models}

\subsection{Theoretical calculations of drift reduction} \label{sec:driftredmodels}

Stochastic transport models of CRs typically include the macroscopic drift of the charged particles in the diffusion tensor that is used to describe the diffusive propagation of cosmic rays due to plasma turbulence. Within the assumption of isotropic particle velocity distribution, the large-scale drift velocity $\mathbf{v}_{d}$ of a charged particle due to gradient and curvature of the magnetic field $\mathbf{B}$ can be written in form
\begin{equation}\label{eq:v_drift}
\mathbf{v}_d=\frac{pv}{3q} \nabla\times \frac{\mathbf{B}}{B^2},
\end{equation}
where $p$, $v$ and $q$ are the moment, speed and charge of the particle, respectively \citep[e.g.][]{Rossi1970_Introductiontophysics}. This can be written as 
\begin{equation}\label{eq:vd_kappaA}
\mathbf{v}_d = \nabla\times\left(\frac{v}{3}{r_L} \hat{\mathbf{e}}_B\right) = \nabla\times\kappa_A \hat{\mathbf{e}}_B,    
\end{equation}
where $r_L=p/(qB)$ is the particle's Larmor radius, $\hat{\mathbf{e}}_B$ is the unit vector along the magnetic field $\mathbf{B}$, and $\kappa_A=\frac{v}{3}{r_L}$ is the so-called antisymmetric diffusion tensor. If we consider this form in conjunction with the diffusive flux term in transport equations, $\nabla\cdot(\kappa\cdot\nabla f)=\partial_i\kappa_{ij}\partial_j f$, where $\kappa$ is the diffusion tensor defined in a coordinate system where $z$-axis is aligned along the magnetic field, $\hat{\mathbf{e}}_z\parallel \hat{\mathbf{e}}_B$, the curl in Equation~(\ref{eq:vd_kappaA}) is equivalent to including antisymmetric off-axis diffusion tensor elements elements $\kappa_{xy}=-\kappa_{yx}=\kappa_A$.

In order to calculate the effect of the turbulence on drifts, \citet{Bieber1997_PerpendicularDiffusionDrift} considered the Taylor-Green-Kubo formalism \citep{Taylor1922_DiffusionContinuousMovements,Green1951_BrownianMotionGas,Kubo1957_Statistical-MechanicalTheoryIrreversible}, where the diffusion tensor is given as
\begin{equation} \label{eq:TKGintegral}
\kappa_{ij}=\int_0^\infty dt \left<v_j(t_0)v_i(t_0+t)\right>
=\int_0^\infty dt R_{ij}(t)
\end{equation}
where $\left<\right>$ represents ensemble average, and 
\begin{equation}
    R_{ij}(t)= \left<v_j(t_0)v_i(t_0+t)\right>
\end{equation}
 is a correlation function that is statistically independent of $t_0$. For a (positively) charged particle in uniform magnetic field, the velocity components are $v_x=v_\perp\cos(\Omega t+\phi_0)$, $v_y=-v_\perp\sin(\Omega t+\phi_0)$ and $v_z=$ constant,
which results in correlation functions
\begin{align}\label{eq:Rxx}
R_{xx}&=R_{yy} =\frac{1}{2}v_\perp^2 \cos(\Omega t) \\
R_{yx}&=-R_{xy}=\frac{1}{2}v_\perp^2 \sin(\Omega t),
\end{align}
where $\Omega=v/r_L$ is the gyrofrequency of the particle. 
\citet{Bieber1997_PerpendicularDiffusionDrift} assumed that because the turbulence disturbs the gyration of the charged particles, the correlation between the velocity components should include a decay term as follows: 
\begin{equation}
    R_{yx}=-R_{xy}=\frac{1}{2}v_\perp^2 \sin(\Omega t)\, \mathrm{e}^{-t/\tau}
\end{equation}
where $\tau$ is the decorrelation timescale of the gyration. Using this expression for the correlation functions, by integration they obtained the diffusion coefficients
\begin{align}
\kappa_{xx}&=\kappa_{yy}=\frac{v r_L}{3}\frac{\Omega\tau}{1+\Omega^2\tau^2}\\
\kappa_{xy}&=-\kappa_{yx} \equiv \kappa_A = \frac{v r_L}{3}\frac{\Omega^2\tau^2}{1+\Omega^2\tau^2} \label{eq:TGK_kappaxy}
\end{align}
where an average over the pitch angle has been carried out. Comparing this form to the definition of $\kappa_A$ in Equation~(\ref{eq:vd_kappaA}), we see that the decorrelation of the particle velocity at time-scale $\tau$ reduces the drift coefficient by a factor
\begin{equation} \label{eq:TGK_fs}
    f_s = \frac{\Omega^2\tau^2}{1+\Omega^2\tau^2},
\end{equation}
termed the drift reduction factor.

The question then is what is the correct timescale for the decorrelation of the particle gyromotion to be used to evaluate $f_s$? \citet{Bieber1997_PerpendicularDiffusionDrift} approached the question by assuming that the gyromotion decorrelates when the particles, following diffusing field lines, have drifted the distance of their gyroradius from their original position across the mean field direction. Following this argument, they arrived at an expression 
\begin{equation} \label{eq:omegatau_BAM}
    \Omega\tau=\frac{2r_L}{(3D_\perp)},
\end{equation}
where $D_\perp$ is the magnetic field line diffusion coefficient. Thus, the drift reduction coefficient depends on the turbulence amplitude, its associated length scales \citep[e.g.][]{Matthaeus1999_CorrelationlengthsUltrascale}, as well as the particle energy and mass-charge ratio.

Using a similar approach, \citet{Engelbrecht2017_TowardGreaterUnderstanding} and \citet{Vandenberg_2021_TurbulentReductionDrifts} considered that the relevant cross-field scale for the decorrelation should be the perpendicular mean free path of the particles, $\lambda_\perp$. They further assumed that the particle cross-field velocity is determined by the random-walk of the field lines, arriving at 
\begin{equation} \label{eq:omegatau_Eng}
    \Omega\tau=\frac{r_L}{\lambda_\perp}\frac{B_0}{dB},
\end{equation}
where $dB^2$ is the variance of the turbulence.

It should be noted that the derivation of Equation~(\ref{eq:TGK_kappaxy}) does not involve large-scale gradients: it is conducted with the assumption of uniform magnetic field. Thus, it does not describe the large-scale drifts or their reduction, rather it describes asymmetric flux of particles due to their disturbed gyration. 

It should also be noted that in the limit $\tau\rightarrow\infty$, $\kappa_A$ cannot be defined by this approach, as the integral in Equation~(\ref{eq:TKGintegral}) is not defined.

\subsection{Test particle simulations}\label{sec:test-particle-simulations}

The drift reduction due to turbulence has been studied by several researchers by means of full-orbit test particle simulations \citep[e.g.][]{Giacalone1999_ParticleDriftsFluctuating, Candia2004_Diffusiondriftcosmic, Minnie2007_SuppressionParticleDrifts, Tautz2012_DriftCoefficientsCharged}. These works typically superpose homogeneous turbulence on a constant background magnetic field, a configuration that is similar to that used in the theoretical models presented in Section~\ref{sec:driftredmodels}. In such a configuration, measuring a macroscopic drift of a particle population cannot be used to analyse drift or its reduction, as the macroscopic drift requires a gradient or curvature in the macroscopic field. Instead, these studies use the approach suggested by \citet{Giacalone1999_ParticleDriftsFluctuating}, where $\kappa_{ij}=\left<v_i\Delta r_j\right>$, with $v_i$ and $\Delta r_j$ the $i$ and $j$ components, respectively, of velocity and spatial displacement normal to the background magnetic field. Only \citet{Minnie2007_SuppressionParticleDrifts} used simulations where the background magnetic field has a gradient, and macroscopic drifts are present. We shall return to these simulations in Section~\ref{sec:discussion}.

For this work, we analyse IMF guiding centre drift by integrating the full equation of motion for charged particles in the the heliosphere via the simulation framework developed by \cite{Dalla2005_Particleaccelerationthree-dimensional} and \citet{Marsh2013_Drift-inducedPerpendicularTransport}. The IMF is formed of a Parker spiral magnetic field superposed with turbulent fluctuations as presented in \citet{Laitinen2023_AnalyticalModelTurbulence}. It should be noted that in the present work, the model is monopolar, in form 
\begin{equation} \label{eq:parker}
    \mathbf{B}= A\, B_0\, \frac{r_0^2}{r^2} \left[\hat{\mathbf{e}}_r - \frac{r}{a} \hat{\mathbf{e}}_\phi\right]
\end{equation}
where $A$ is the sign of the magnetic field, $a=v_{sw}/(\Omega_\odot \sin\theta)$, with $v_{sw}$ and $\Omega_\odot$ the solar wind speed and solar rotation rate respectively, $B_0\approx B(r_0)$ when $r_0<<a$, with $r_0$ a reference location close to the Sun, and $\theta$ the colatitude. Thus, our present work does not include the  effects of the heliospheric current sheet \citep{Battarbee2017_SolarEnergeticParticle, Battarbee2018_ModelingSolarEnergetic}. We denote the polarity of the IMF with B+ for IMF pointing away from the Sun, with $A=1$, and B- for IMF pointing towards the Sun, with $A=-1$. From the equation of the magnetic drifts, Eq.~(\ref{eq:v_drift}), it is clear that the drift velocity changes its direction with different IMF polarities. We note that the model used in the present work also does not include the convectional or turbulent electric field, thus the corotation drift and deceleration of CRs are not present in the simulations.

For the turbulence in our model, we use the heliospheric analytic 2D-slab composite turbulence model model described in \citet{Laitinen2023_AnalyticalModelTurbulence}.  The model consists of 2D turbulence component which is transverse with respect to the Parker spiral, and has the fluctuating magnetic field $\delta\mathbf{B}$ normal to the Parker spiral everywhere. The 2D component is complemented with weaker slab-like component, which is dominated by radial slab modes close to the Sun, and azimuthal at larger distances. The turbulence is realised as a sum of Fourier modes following the approach of \citet{Giacalone1999_TransportCosmicRays}, with 128 each of slab and 2D-mode waves which are used to calculate the magnetic field at the particle location. Note that this approach differs from the approaches where the turbulent magnetic field is pre-calculated on a 3-dimensional grid using a Fourier transform: in such case, the simulation domain is limited by the largest scales included in the turbulence spectra. We refer to simulations performed with this model as "turbulence" model throughout the manuscript.

We compare results from the turbulence model to heliospheric particle simulations without superposed turbulence. To do that, we run simulations using the approach presented in \citet{Marsh2013_Drift-inducedPerpendicularTransport}, where the particles are traced with full 3D test particle simulations in Parker spiral geometry with $\mathbf{B}$ given by Equation~(\ref{eq:parker}) without turbulent field-line meandering. In these simulations, the effect of the turbulence on the particles is modelled as scattering events where the particle's velocity vector is scattered at random times, parameterised by a constant parallel scattering mean free path, $\lambda_\parallel$ \citep{Marsh2013_Drift-inducedPerpendicularTransport}. We refer these simulations as the "scatter" model. Note that cross-field scattering is not explicitly implemented in the scatter model: the only cross-field motion is caused by large-scale drifts and the random-walk of the particle gyrocentre as the velocity vector is randomised.
For the scatter and turbulence approaches to be comparable, we have used a parallel mean free path according to standard quasi-linear theory (SQLT) \citep[e.g.][]{Jokipii1966_Cosmic-RayPropagationI} derived from the turbulence model parameters used in the turbulence runs. As the obtained SQLT mean free path varies with radial distance from the Sun, we average its value over distances from $2 r_\odot$ to 1~au.

In this work, we simulate energetic protons at different proton energies and interplanetary space conditions. For each parameter set, we simulate 1,000 particles in 100 different turbulence realisations, with the turbulence mode phases and polarisations different in each realisation, thus a total of 100,000 particles for each simulation set. The particles are injected at $2 r_\odot$ heliocentric distance at the solar equator, with isotropic velocity distribution, and their propagation in the heliosphere is traced for 48~hr. The parameters of the eight simulation sets are given in Table~\ref{tab:drifts}, with each set including both a turbulence and a scatter simulation run (column 5). We simulate energetic protons from non-relativistic to relativistic energies, at 10, 100 and 1000~MeV (column 2 in Table ~\ref{tab:drifts}), for the two polarities of the monopolar IMF (column 3). The relative turbulence amplitude at 1~au, used in the turbulence simulations, is given in column~7 of Table~\ref{tab:drifts}, with the corresponding scattering mean free path, used in the scattering simulations, given in column~6. Other turbulence parameters are as in \citet{Laitinen2023_Solarenergeticparticle}, briefly described in Appendix~\ref{app:turbparams}.

\subsection{Determining the latitudinal drift velocity}

Previous studies have derived drift coefficients in a uniform background magnetic field with homogeneous turbulence: in this type of configuration the coefficients do not depend on the particle location, thus the analysis methods can average over particles over the entire simulation volume. In our heliospheric configuration, however, use of such methods is not possible: turbulence characteristics, and thus also the particle transport parameters, depend significantly on heliospheric location \citep[e.g.][]{Chhiber2017_Cosmic-RayDiffusionCoefficients}. Drift velocities also vary significantly with particle location \citep[e.g.][]{Dalla2013_SolarEnergeticParticle}. Thus, the coefficients obtained by averaging over the simulation volume (by tracing particles through the entire heliosphere) would not represent the coefficients at any given location, but an average value over the whole heliosphere. To alleviate this issue, and produce coefficients for our turbulence and scatter particle simulations that can be compared with each other, we have developed a new methodology to analyse the drift experienced by the particles.

\begin{figure}[t]
    \includegraphics[width=\columnwidth]{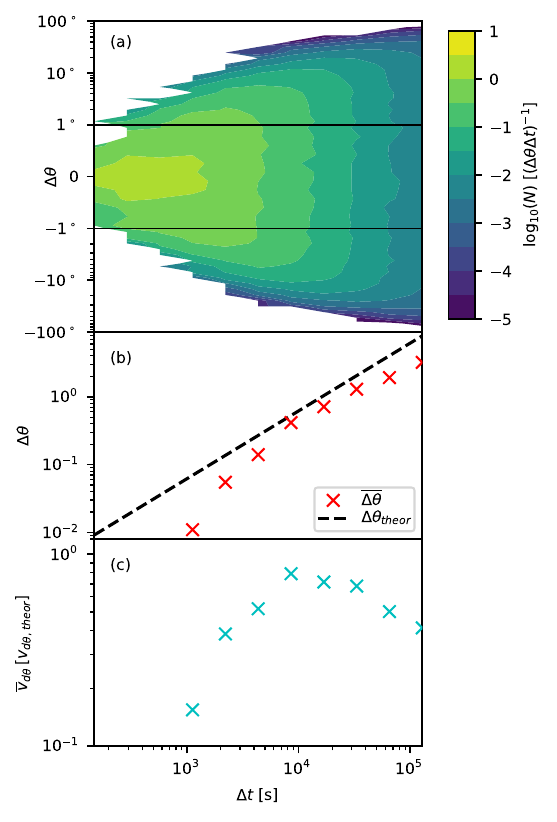}
    \caption{ Panel (a): The distribution of the latitudinal displacements $\Delta\theta$ between the first and last 1-au crossings of 100 MeV protons (simulation set~4) as a function of the time $\Delta t$ between the first and last crossings, ensemble-averaged over all turbulence realisations (Note that the interval of $\Delta\theta \in [-1,1]$ on the vertical axis is linear). In panel (b), the red crosses in the top panel show $\Deltatheta$, and the dashed line the theoretical $\Delta\theta$ due to guiding centre drifts (see Appendix~\ref{app:drift1au}). Panel (c) shows the drift velocity $\vtheta$ in units of the theoretical velocity.  \label{fig:dtheta_dist}}
\end{figure}

\begin{table*}[]
    \caption{The parameters of the simulation runs. The first column gives the set identifier for the runs, with the rows for the  scatter and turbulence simulation runs within the sets indicated in column~5. The second column shows the proton energy, third the magnetic field polarity and fourth the theoretical $v_{d\theta\,theor}$, as calculated in Appendix~\ref{app:drift1au}. The sixth and seventh columns give the parallel scattering mean free path and turbulence variance at 1~au heliocentric distance, used for the scatter and turbulence simulations, respectively. The eighth and ninth columns give $\left<\vtheta\right>_r$ and the uncertainty $\sigma$, obtained from the distributions shown in Figure~\ref{fig:driftstats}.}
    \label{tab:drifts}
    \centering
\centering                                      
\begin{tabular}{crlrrrrclrrr}
\hline\hline
Set & $E_k$ [MeV] & Pol & \multicolumn{2}{c}{$v_{d\theta\,theor}\,\,[\mathrm{km\, s^{-1]}}]$} & sim & $\lambda_\parallel$ [au] & $dB^2/B^2$ & \multicolumn{2}{c}{$\overline{\left<\vthetaPlain\right>}_r\,\,[\mathrm{km\, s^{-1]}}]$} &   \multicolumn{2}{c}{$\sigma\,\, [\mathrm{km\, s^{-1]}}]$}\\
\hline
1 & 10 & B+ & 17. & & scat &0.37 &  -- &14. & & 0.38 & \\
 & &  &  & & turb & -- & 0.60 & 3.5 & &19. & \\
2 & 10 & B- & 17. & & scat &0.37 &  -- &-14. & & 0.39 & \\
 & &  &  & & turb & -- & 0.60 & -4.7 & &17. & \\
3 & 100 & B+ & 160. & & scat &1.6 &  -- &110. & & 2.2 & \\
 & &  &  & & turb & -- & 0.20 & 110. & &81. & \\
4 & 100 & B+ & 160. & & scat &0.55 &  -- &120. & & 2.3 & \\
 & &  &  & & turb & -- & 0.60 & 82. & &64. & \\
5 & 100 & B+ & 160. & & scat &0.17 &  -- &130. & & 3.3 & \\
 & &  &  & & turb & -- & 2.00 & 40. & &58. & \\
6 & 100 & B- & 160. & & scat &0.55 &  -- &-120. & & 2.6 & \\
 & &  &  & & turb & -- & 0.60 & -82. & &53. & \\
7 & 1000 & B+ & 1300. & & scat &0.87 &  -- &840. & & 18. & \\
 & &  &  & & turb & -- & 0.60 & 780. & &300. & \\
8 & 1000 & B- & 1300. & & scat &0.87 &  -- &-840. & & 18. & \\
 & &  &  & & turb & -- & 0.60 & -800. & &300. & \\ \hline
\end{tabular}
\end{table*}

Our method is based on the fundamental difference between the effects of the two distinct physical processes (turbulence and drift) on the particle distribution. We initialise our particles around the solar equator, at $\theta=\pi/2$. Our turbulence model is symmetric in latitude, thus any asymmetry in the distribution in latitude with respect to the solar equator cannot be caused by the stochastic turbulence: without drift, the mean of the particle distribution in colatitude would remain at $\theta=\pi/2$. On the other hand, in a unipolar field drift in heliolatitude causes particles to propagate systematically either northwards or southwards depending on the magnetic field polarity ($A$ in Equation~(\ref{eq:parker}) \citep{Dalla2013_SolarEnergeticParticle}. Thus, we consider the broadening of the particle distribution to be caused by turbulence, and any changes in the mean latitude of the particle distribution to be caused by large-scale drifts.

To analyse the latitudinal propagation of a particle, we consider its first and last crossings of the 1~au sphere, taking place at times $t_f$ and $t_l$, respectively. The change in colatitude between the two crossing is
\begin{equation}
 \Delta \theta = \theta_l - \theta_f      
\end{equation}
and the time interval between the crossings is
\begin{equation}
 \Delta t = t_l - t_f.   
\end{equation}
The  latitudinal velocity of the particle averaged between the first and last 1~au sphere crossings, $v_{d\theta}$, can be then defined as
\begin{equation} \label{eq:vdtheta_definition}
v_{d\theta}=r_e\frac{\Delta\theta}{\Delta t},     
\end{equation}
where $r_e=1\,\mathrm{au}$. It should be noted that $v_{d\theta}$ does not represent the 1~au value of the latitudinal velocity: rather, it is the average latitudinal velocity of the particle over its entire trajectory in the heliosphere between $t_f$ and $t_l$.

We use the interval between the first and the last crossings instead of interval between consecutive crossings in our definitions to avoid giving larger statistical weight to individual particles that cross the 1-au sphere multiple times. Our method also ignores the initial propagation of the particles in the near-Sun region where the stochastic spreading of the particle population is strong: as discussed in \citet{Laitinen2023_AnalyticalModelTurbulence}, the field-line meandering in such an environment can cause large local deviations. We further exclude particles for which $\Delta t<100$~s, large compared to $\Omega=0.49\; \mathrm{s}^{-1}$ at 1~au (for non-relativistic particles), to avoid effects of the Larmor radius of the particles on $\Delta \theta$.

Figure~\ref{fig:dtheta_dist}~(a) shows the distribution of $\Delta\theta$ values for 100~MeV protons (filled contours). Here, the entire population of 100,000 protons over the 100 turbulence realisations is used. The particles with small $\Delta t$ have returned to 1~au soon after their initial crossing, and have  subsequently escaped to further distances without returning. Particles with larger $\Delta t$ display a wider distribution in $\Delta\theta$. This is due to the particles decoupling from their original meandering field line to different field lines at later times. This stochastic process can result in a particle's final crossing of the 1-au sphere to be up to $90^\circ$ from their initial crossing colatitude.

\begin{figure*}[t]
    \centering
    \includegraphics{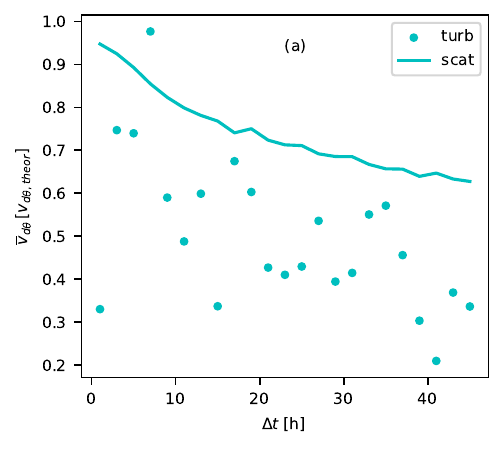}
    \includegraphics{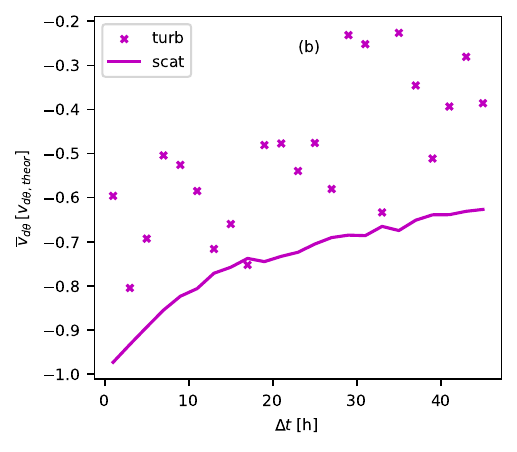}
    \caption{The drift velocity $v_{d\theta}$ (in units of the theoretical drift velocity, Equation~\ref{eq:theor_drift_vel})) of 100~MeV protons in moderate turbulence for (a) B+ -polarity (simulation set 4) and (b) B- -polarity (simulation set 6) protons as a function of time interval $\Delta t$ between the first and last 1-au crossing.}
    \label{fig:dtheta_vs_t}
\end{figure*}

To investigate the temporal evolution of $\Delta\theta$, we divide the $\Delta t$ axis into bins and calculate median values of $\Delta\theta$ over the particle population in each bin,  indicated as $\overline{\Delta\theta}(\Delta t)$. We show $\Deltatheta$ as function of $\Delta t$ in Figure~\ref{fig:dtheta_dist}~(b) with red crosses. As can be seen, $\overline{\Delta\theta}$ increases systematically, demonstrating a macroscopic, systematic drift of the particle population in time.

We compare the drift in Figure~\ref{fig:dtheta_dist}~(b) to the theoretical drift. The theoretical drift velocity (Equation~(\ref{eq:theor_drift_vel})) is given in column~4 of Table~\ref{tab:drifts}, following the derivation presented in Appendix~\ref{app:drift1au}. The dashed black curve in Figure~\ref{fig:dtheta_dist} shows change in colatitude of a particle obeying the theoretical drift as
\begin{equation}
    \Delta\theta_{theor}=v_{d\theta\,theor} \Delta t.
\end{equation}
As can be seen, the $\overline{\Delta\theta}$ follows the trend of the theoretical prediction well, however at slightly lower values.

To further analyse the drift in velocity units, we calculate the median drift speed $\vtheta(\Delta t)$
and present it in units of $v_{d\theta\,theor}$ in Figure~\ref{fig:dtheta_dist}~(c). As can be seen, $\vtheta(\Delta t)$ is of the same order as $v_{d\theta\,theor}$, but lower in magnitude, reaching its maximum value of $0.85\; v_{d\theta\,theor}$ at $\Delta t \approx 10,000$ seconds.

It should be emphasised that the wide, temporally widening extent of the $\Delta\theta$ distribution seen in Figure~\ref{fig:dtheta_dist}~(a) cannot be considered as an uncertainty of $\Deltatheta$. The extent of the distribution arises from the  decoupling  of the particles from the stochastically meandering field lines: it is a measurable quantity that describes the turbulence-induced stochastic particle propagation. The wide extent is a result of a physical process that is distinct from large-scale systematic drifts which are caused by large-scale gradients and curvature of the Parker spiral.

In the next section, we use our new methodology to investigate the dependence of the drift velocity on the particle energy and turbulence amplitude, as well as its temporal evolution and variation across different turbulence realisations. For the latter, it is useful to condider the distribution of the median drift velocities over the 100 realisations. The median drift velocity over realisation $r$, and over all $\Delta t$ values, is indicated as $\left<\vthetaPlain\right>_r$.

\section{Results}\label{sec:results}

\subsection{Drifts from test particle simulations}

We first investigate the temporal evolution of $\vtheta(\Delta t)$
in Figure~\ref{fig:dtheta_vs_t} for the simulation sets~4 and~6, in panels~(a) and~(b) respectively, in units of the theoretical drift velocity. In the Figure, the cyan and magenta symbols depict $\vtheta(\Delta t)$ for the turbulence simulations, and the solid cyan curve the same for the scatter simulation sets~4 and~6. Note that the binning in $\Delta t$ is linear unlike in Figure~\ref{fig:dtheta_dist}~(c) where we used logarithmic binning.

As can be seen, both the turbulence and scatter simulations attain $\vtheta(\Delta t)$ values with the same sign as the polarity used in the simulation set. The drift from the scatter simulations is initially of the same order as $\vthetaTheor$ (i.e., values 1 and $-1$ in Figure~\ref{fig:dtheta_vs_t}~(a) and~(b), respectively), however it decreases for larger $\Delta t$. This is caused by the particle propagation not being limited to the vicinity of 1~au where the theoretical drift velocity is calculated. As the particles propagate and spread into the inner and outer heliosphere before returning to 1~au, they sample heliocentric distances where the angular drift velocity is smaller than at 1~au (see Figure~\ref{fig:vdtheta_theor}~(b)). As a result, their integrated drift velocity between the first and last crossings are smaller than 1~au values. This effect is evident in that for the particles with larger $\Delta t$, the deviation of the drift speed from the theoretical value at 1~au is progressively larger. For further details, see Appendix~\ref{app:drift1au}).

The $\vtheta(\Delta t)$ from turbulent simulations is smaller in magnitude than the one from scatter simulations, and likewise is smaller for larger $\Delta t$. Thus, it appears that some reduction of drift is present in the turbulence simulations as compared to the scatter simulations. There is considerable statistical fluctuations, due to different travel histories of the particles within the simulations.

\begin{figure}[ht]
\centering
\includegraphics[width=\columnwidth]{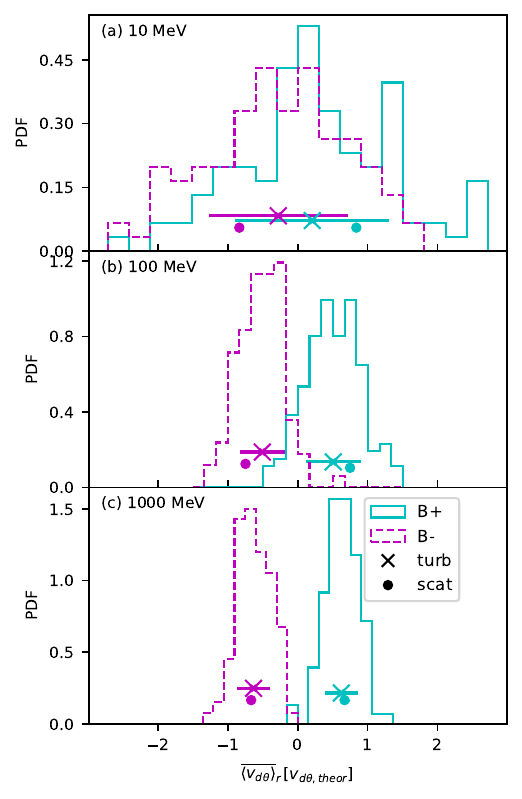}
\caption{\label{fig:driftstats} Probability density of $\left<\vthetaPlain\right>_r$, the medians of the $\vthetaPlain$ of the 100 different turbulence realisations, for $dB^2/B^2=0.6$ at 1~au for (a) 10 MeV, (b) 100 MeV and (d) 1000 MeV protons, in units of $\vthetaTheor$. The cyan and magenta curves show $\left<\vthetaPlain\right>_r$  distribution for B+ and B- polarities, respectively. The crosses and horizontal error bars show the median and uncertainty for the turbulence simulations, and the filled circles show the median for the scatter simulations.}
\end{figure}

In order to evaluate the effect of different turbulence realisations on the drift velocity obtained from our simulations, we show the distributions of $\left<\vthetaPlain\right>_r$ in Figure~\ref{fig:driftstats} for the three proton energies used in this study for the moderate turbulence case of $dB^2/B^2=0.6$ at 1~au, with the positive (B+) and negative (B-) magnetic polarity simulations shown by cyan and magenta symbols and lines, respectively. The crosses and horizontal error bars show the ensemble median $\overline{\left<\vthetaPlain\right>}_r$ and the standard deviation $\sigma$ over the distribution of 100 $\left<\vthetaPlain\right>_r$ values for the turbulence realisations. The values are given in Table~\ref{tab:drifts} in the last two columns on lines denoted "turb".

For comparison, we show in Figure~\ref{fig:driftstats} the $\overline{\left<\vthetaPlain\right>}_r$ for the scatter simulations with filled circle symbols. The standard deviations of the scatter simulations are smaller than the symbol size. The value of $\vthetaScat$ and the corresponding standard deviation are shown in the last two columns of Table~\ref{tab:drifts} on lines denoted "scat". 

As we can see in Figure~\ref{fig:driftstats}, the distributions of median drifts for the B+ and B- polarities have different signs, and for the higher energies the $\vthetaTurb$ distributions for the different polarities are clearly distinct. For the 10~MeV protons, the distributions do overlap significantly, however the median values of the distributions differ and are of different sign, as expected. In addition, we performed a Kolmogorov-Smirnoff test which confirmed that the B+ and B- distributions are statistically different even for the 10~MeV protons. In all the cases depicted in Figure~\ref{fig:driftstats} and in Table~\ref{tab:drifts}, the drift obtained with the scatter simulations is larger than $\overline{\left<\vthetaPlain\right>}_r$.

The theoretical drift velocity, shown in column~4 of Table~\ref{tab:drifts}, is of similar order as the scatter simulation drift, however there are some differences. This is due to the fact that the theoretical value is valid only for particles at 1~au, whereas the simulated particles propagate in the IMF where the drift rate varies with heliocentric distance (see Figure~\ref{fig:vdtheta_theor}~(b)) and colatitude.

\subsection{Drift reduction factor}

\begin{figure*}[ht]
    \centering
    \includegraphics{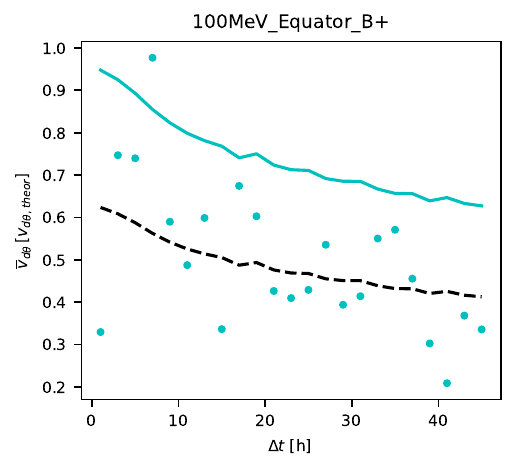}
    \includegraphics{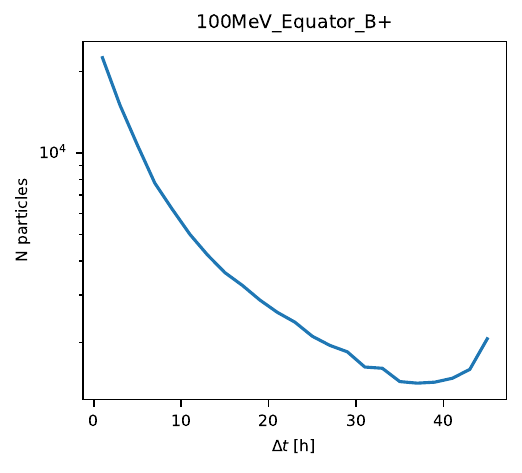}
    \caption{(a) $v_{d\theta}$, in units of $\vthetaTheor$, of 100 MeV protons (simulation sets 4) as a function of $\Delta t$ with symbols showing the turbulence simulations and cyan line the scatter simulations. The dashed black line shows the result of fitting Equation~(\ref{eq:fs_fit}). (b) The number of particles corresponding the $v_{d\theta,scat}$ in (a), used as weights for fitting Equation~(\ref{eq:fs_fit}).}
    \label{fig:dtheta_vs_t_fit}
\end{figure*}

To evaluate the effect of the turbulence on drifts, we compare $\vtheta(\Delta t)$ obtained with turbulence simulations with the drift expected without turbulence. As discussed above, $\vthetaTheor$ calculated in Appendix~\ref{app:drift1au} is limited to a specific heliospheric location. For this reason, we choose the drift from the scatter simulations as the reference drift without turbulence. This assumption can be justified by considering the form of Equation~(\ref{eq:TGK_fs}) which approaches unity for $\Omega\tau\gg 1$. For the scatter model, we can consider the decorrelation time scale $\tau$ to be equal to the parallel scattering time $t_{\mathrm{scat}}=\lambda_\parallel/v$ \citep{Marsh2013_Drift-inducedPerpendicularTransport}, which for all of our parameters is much larger than $1/\Omega$. 

Under this assumption, we obtain the drift reduction coefficient by fitting the time-dependent $\vthetaTurb$, as shown in Figure~\ref{fig:dtheta_vs_t} as 
\begin{equation} \label{eq:fs_fit}
    \vthetaTurb = f_s \,\vthetaScat
\end{equation}
where $f_s$ is the drift reduction factor. We note that the number of crossings also depends on $\Delta t$, thus we fit Equation~(\ref{eq:fs_fit}) weighting the turbulence simulation points with the number of particles. We perform the fit using the curve\_fit function of the python scipy package \citep{2020SciPy-NMeth}. We show an example fit in Figure~\ref{fig:dtheta_vs_t_fit}, where panel~(a) shows the turbulence and scattering simulation $\vtheta(\Delta t)$ with cyan circles and line for 100~MeV protons (simulation set~4), respectively, and the fit to Equation~(\ref{eq:fs_fit}) with black dashed curve. In panel~(b) the blue curve shows the number of particles used as weights for the fitting. As can be seen in panel (b), the number of 100~MeV protons used in the fitting decays as a function of $\Delta t$, thus, due to weighting, the low-$\Delta t$ $\vtheta(\Delta t)$ affect the fit more strongly than the high-$\Delta t$ $\vtheta(\Delta t)$.

\begin{figure}[ht]
 \includegraphics[width=\columnwidth]{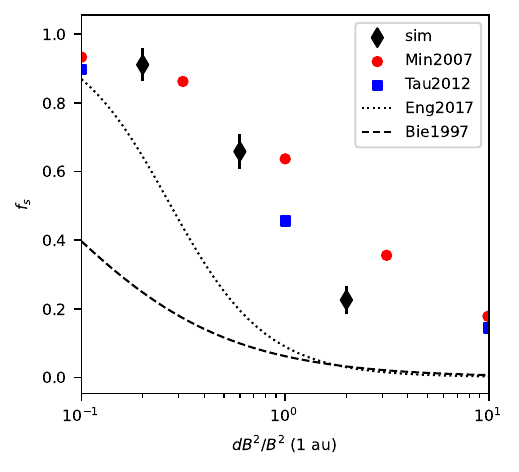}
  \caption{Drift reduction factor for 100 MeV protons as a function of relative turbulence variance, from our simulations (black diamonds) and for the models by \citet{Bieber1997_PerpendicularDiffusionDrift} (dashed curves) and \cite{Engelbrecht2017_TowardGreaterUnderstanding} (dotted curves). The red circles and blue squares are from the \citet{Minnie2007_SuppressionParticleDrifts} and \citet{Tautz2012_DriftCoefficientsCharged} simulations, respectively, with $r_L/\lambda_{b\parallel}=0.1$, corresponding to proton energy 227~MeV.}\label{fig:driftred_dB2B2}
\end{figure}

\begin{figure}[ht]
 \includegraphics[width=\columnwidth]{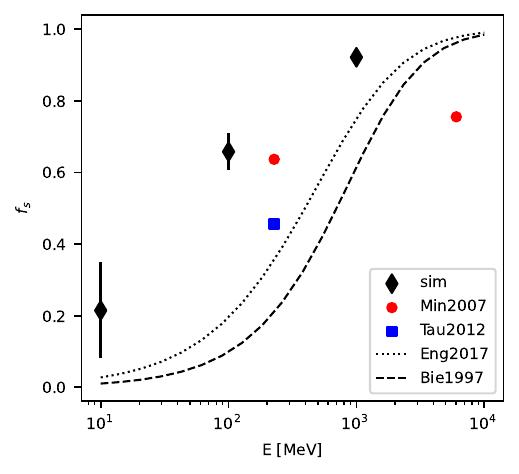} 
  \caption{Drift reduction factor for turbulence amplitude $dB^2/B^2=0.6$ at 1 au, as a function of proton energy, from our simulations (black diamonds) and for \citet{Bieber1997_PerpendicularDiffusionDrift} (dashed curves) and \cite{Engelbrecht2017_TowardGreaterUnderstanding} (dotted curves) theoretical models. The red circles and blue squares are from the  \citet{Minnie2007_SuppressionParticleDrifts} and \citet{Tautz2012_DriftCoefficientsCharged} simulations, respectively, with $dB^2/B^2=1$.}\label{fig:driftred_Ekin}
\end{figure}

The drift reduction factors and their standard deviations obtained from the fitting procedure are shown in the 5th and 6th columns of Table~\ref{tab:driftred}. We demonstrate the dependence of $f_s$ on the relative amplitude of turbulence and proton energy in Figures~\ref{fig:driftred_dB2B2} and~\ref{fig:driftred_Ekin}, respectively, for the B+ polarity.

For comparison, we have also calculated the theoretical reduction factors from the models by \citet{Bieber1997_PerpendicularDiffusionDrift} and \citet{Engelbrecht2017_TowardGreaterUnderstanding} using Equations~(\ref{eq:TGK_fs})-(\ref{eq:omegatau_Eng}). The field line diffusion coefficient and particle cross-field diffusion coefficient required for these were calculated using the random ballistic decorrelation (RBD) approach by \citet{Ghilea2011_MagneticFieldLine} and \citet{Ruffolo2012_RandomBallisticInterpretation}, respectively\footnote{Note that the reduction factor of \citet{Engelbrecht2017_TowardGreaterUnderstanding} depends strongly on the choice of the theory used to calculate the perpendicular diffusion coefficient, see \citet{Vandenberg_2021_TurbulentReductionDrifts}.}. These values are shown in columns~7 and~8 of Table~\ref{tab:driftred}. A comparison of these drift reduction factors are also shown in Figures~\ref{fig:driftred_dB2B2} and~\ref{fig:driftred_Ekin}. As can be seen, our reduction factor is considerably larger than that predicted by the theoretical approach, except for high energies and low turbulence amplitudes.

We note that care should be taken in comparing simulations with the TGK approach, as the simulation timescales should be much larger than the decorrelation timescale $\tau$ for TGK to be applicable. We can estimate the validity of the TGK approach by using the theoretical $f_s$ values presented in Table~\ref{tab:driftred}, where all $f_s$ values are at or below 0.90. Using Equation~(\ref{eq:TGK_fs}) $f_s<0.90$ corresponds to $\tau\Omega>3$, which, with the (non-relativistic) gyrofrequency at 1~au being $\Omega=0.49\;\mathrm{s}^{-1}$ for our simulations corresponds to $\tau\approx 6\;\mathrm{s}$. As we exclude particles with $\Delta t<100\;\mathrm{s}$ from our analysis, we can conclude that for all our simulations $\Delta t\gg \tau$, and the TGK approach is valid.

We show also the drift reduction coefficients obtained from full-orbit particle simulations by \citet{Minnie2007_SuppressionParticleDrifts} and \citet{Tautz2012_DriftCoefficientsCharged} who give the energy of particles in terms of $r_L/l_{b\parallel}$, where $l_{b\parallel}$ is the largest scale of the inertial scale of the slab spectrum, which for our turbulence model is $l_{b\parallel}=0.27$~au at $r=1$~au \citep{Laitinen2023_AnalyticalModelTurbulence}. \citet{Minnie2007_SuppressionParticleDrifts} used values $r_L/l_{b\parallel}=0.1$ and 1, which correspond to proton energies 227 and 6040 MeV, whereas \citet{Tautz2012_DriftCoefficientsCharged} only used $r_L/l_{b\parallel}=0.1$. We give $r_L/l_{b\parallel}$ for our simulations in column~3 of Table~\ref{tab:driftred}. As can be seen in Figure~\ref{fig:driftred_dB2B2}, the dependence of our drift reduction factor  on the turbulence amplitude is similar to that obtained by \citet{Minnie2007_SuppressionParticleDrifts} and \citet{Tautz2012_DriftCoefficientsCharged}. The dependence of our drift on energy differs more significantly particularly from the \citet{Tautz2012_DriftCoefficientsCharged} result (Figure~\ref{fig:driftred_Ekin}), however this can be explained by the fact that the relative turbulence variance $dB^2/B^2$ is~1 in their work, whereas our simulations in Figure~\ref{fig:driftred_Ekin} have $dB^2/B^2=0.6$ at 1~au. By estimating the effect of the turbulence amplitude on $f_s$ in Figure~\ref{fig:driftred_dB2B2}, it is clear that reducing the turbulence amplitude in the Minnie and Tautz simulations would increase $f_s$ considerably. 

Thus, we can conclude that our simulations are well in line with the previous simulation work. It should be noted though that our, \citet{Minnie2007_SuppressionParticleDrifts} and \citet{Tautz2012_DriftCoefficientsCharged} simulations are not directly comparable, as the particle and turbulence parameters in the studies are different, and we use heliospheric magnetic field whereas the \citet{Minnie2007_SuppressionParticleDrifts} and \citet{Tautz2012_DriftCoefficientsCharged} use constant background magnetic field. We discuss these issues further in Section~\ref{sec:discussion}.

\begin{table}[]
\caption{Drift reduction factors and their standard deviations obtained from the simulations (columns 3 and 4), and calculated with the turbulence and particle parameters using the \citet[Bie1997]{Bieber1997_PerpendicularDiffusionDrift} (column 5) and \citet[Eng2017]{Engelbrecht2017_TowardGreaterUnderstanding} (column 6) models. The first column identifies the simulation set used in Table~\ref{tab:drifts}, the second the ratio of the particle's Larmor radius and the breakpoint scale of the slab component of the turbulence, $r_L/l_{b\parallel}$.
All quantities are calculated at 1~au heliocentric distance.
}
\label{tab:driftred}
\centering                                      
\begin{tabular}{clllccc}
\hline\hline
Set & $r_L/l_{b\parallel}$ & $f_s$ & $\sigma_{fs}$ & $f_s$ Bie1997 & $f_s$ Eng2017 \\
\hline
1 & 0.0199 & 0.21 & 0.13 & 0.010 & 0.045 \\
2 & 0.0199 & 0.26 & 0.13 & 0.010 & 0.045 \\
3 & 0.0643 & 0.91 & 0.048 & 0.25 & 0.90 \\
4 & 0.0643 & 0.66 & 0.052 & 0.099 & 0.29 \\
5 & 0.0643 & 0.23 & 0.042 & 0.032 & 0.015 \\
6 & 0.0643 & 0.71 & 0.03 & 0.099 & 0.29 \\
7 & 0.245 & 0.92 & 0.019 & 0.61 & 0.83 \\
8 & 0.245 & 0.92 & 0.021 & 0.61 & 0.83 \\ \hline
\end{tabular}
\end{table}

\section{Discussion}\label{sec:discussion}

In this work we have investigated the effect of turbulence on large-scale guiding centre drifts of SEPs by using full-orbit test particle simulations. We presented a methodology to evaluate the drift in latitude in turbulent heliospheric simulations, and compared it with simulations not including turbulence. We derived the reduction factor of large-scale drifts in heliospheric magnetic field by comparing two sets of simulations: The turbulent simulations where the Parker spiral magnetic field was overlaid by composite turbulent magnetic field as described by \citet{Laitinen2023_AnalyticalModelTurbulence}, and scatter simulations where the particles propagate in the Parker spiral and experience ad-hoc scattering  \citep{Marsh2013_Drift-inducedPerpendicularTransport}. Our results indicate that turbulence does reduce the large-scale drifts of cosmic rays in the inner heliosphere, particularly at lower energies. The 10~MeV proton drift was reduced to 21\% of the non-turbulent drifts ($f_s=0.21$) for moderate interplanetary turbulence conditions (B+polarity), whereas for higher energies, 100 and 1000~MeV, the drift reduction coefficients are  0.66 and 0.92, respectively. The drift reduction also depended strongly on the relative amplitude of turbulence: increasing turbulence from the level of $dB^2/B^2=0.2$ to~2 (as defined at 1~au heliocentric distance), the reduction factor $f_s$ for 100~MeV protons decreased from 0.91 to 0.23.

We also compared our results to earlier work on drift reduction. Before delving into details of the comparison, it should be noted that our work differs from previous work in two important aspects. First, our magnetic field is not constant but varies in colatitude and in radial distance. Thus, the theoretical drift velocity of the particles varies depending on the location of the particles. Secondly, the turbulence model parameters, such as the spectral shape, amplitude and the turbulence geometry, differs between our and previous work. 

Let us first consider the spatial variation of the magnetic field. Most previous modelling work concentrated on investigating the drift reduction using full-orbit simulations of particles in homogeneous, constant-background magnetic field  \citep[e.g.][]{Giacalone1999_ParticleDriftsFluctuating, Candia2004_Diffusiondriftcosmic, Tautz2012_DriftCoefficientsCharged}. In this configuration, there are no macroscopic drifts as there are no macroscopic gradients and curvatures in the magnetic field. Rather than analysing a macroscopic drift of particles across the mean field, these works have investigated the quantity $\left<v_i\Delta r_j\right>$, where $v_j$ and $\Delta r_i$ are velocity and spatial deviation components $i$ and $j$ normal to the background magnetic field, respectively, which equals the diffusion tensor element $\kappa_{ij}$, as given by \citet{Giacalone1999_ParticleDriftsFluctuating}. It is not clear to the authors of this paper how applicable a drift reduction coefficient calculated from a theoretical starting point without macroscopic drifts is when macroscopic drifts are present.

The only work known to us that does investigate drifts using full-orbit simulations in non-homogeneous magnetic field is \citet{Minnie2007_SuppressionParticleDrifts}, who introduce a gradient normal to the mean magnetic field. With such a configuration, they were able to quantify the drift velocity and its dependence on turbulence amplitude and particle energy. Unfortunately, they only investigate two particle energies, parameterised by the ratio of the particle Larmor radius and the slab breakpoint scale, $r_L/\lambda_{b\parallel}$. For those values, at $dB^2/B^2=1$, their drift reduction coefficients as defined from the drift velocity in their simulations are 0.33 and 0.85 for $r_L/\lambda_{b\parallel}=0.1$ and~1, respectively. 
It is interesting to note that inclusion of the gradient in \citet{Minnie2007_SuppressionParticleDrifts} does not seem to affect the drift coefficient calculated with $\left<v_i\Delta r_j\right>$, as shown in their Figures~4 and~5.

The second issue to note is the differences in turbulence spectra used by the various simulation studies. As discussed above, the ratio between the parallel and perpendicular scales, as well as spectral shapes, are different in our and \citet{Minnie2007_SuppressionParticleDrifts} studies. Drift reduction simulation studies have used a variety of turbulence parameters and models: \citet{Giacalone1999_ParticleDriftsFluctuating} and \citet{Candia2004_Diffusiondriftcosmic} consider isotropic turbulence, whereas \citet{Tautz2012_DriftCoefficientsCharged} investigates the drift reduction for slab, 2D, composite and isotropic turbulence, however for $l_{sl}=l_{2D}$ and different slab/2D energy ratio for their composite model than \citet{Minnie2007_SuppressionParticleDrifts} and our study. Further, they only sample different energies for the isotropic case. The lack of full sampling of the parameter space is a common shortcoming of all these studies, most likely due to the simulations being very time-consuming. In particular the energy-dependence of the drift reduction has not been well-covered by the simulation studies for the case of the composite turbulence, as can be seen in Figure~\ref{fig:driftred_Ekin}. 

Thus, it is very difficult to draw conclusions from the earlier studies of drift reduction, particularly when it comes to the energy dependence of the drift reduction. \citet{Burger2010_ReductionDriftEffects} derive a parametrised form of the drift reduction coefficient to be used in GCR modulation models using the \citet{Minnie2007_SuppressionParticleDrifts} results, however they are limited with the two energies included in that study. The theoretical model by \citet{Engelbrecht2017_TowardGreaterUnderstanding} shows an improved fit to the \citet{Minnie2007_SuppressionParticleDrifts} results, as compared to \citet{Bieber1997_PerpendicularDiffusionDrift}, however as shown by \citet{Vandenberg_2021_TurbulentReductionDrifts}, the drift coefficient depends very strongly on which theory is used to derive the perpendicular diffusion coefficient, used in the \citet{Engelbrecht2017_TowardGreaterUnderstanding} approach. 

A recent study by \citet{Engelbrecht2022_TheoryCosmicRay} showed that the RBD model used in this paper results  in larger particle perpendicular mean free paths in the heliosphere than the field-line random walk \citep[FLRW][]{Jokipii1966_Cosmic-RayPropagationI} and non-linear guiding center \citep[NLGC][]{Matthaeus2003_NonlinearCollisionlessPerpendicular} theories: thus our choice to use RBD for the drift reduction coefficient with the \citet{Engelbrecht2017_TowardGreaterUnderstanding} approach (Equation~(\ref{eq:omegatau_Eng})) would result in smaller $f_s$. However, comparison of the relative magnitudes of the RBD, FLRW and NLGC perpendicular mean free paths is not trivial: in \citet{Matthaeus2003_NonlinearCollisionlessPerpendicular}, the FLRW  mean free path was than the NLGC one, and \citet{Ruffolo2012_RandomBallisticInterpretation} found NLGC mean free path to be larger than the RBD one, both unlike similar comparisons in   \citet{Engelbrecht2022_TheoryCosmicRay}. We believe this may be related to other parameters used in the above mentioned studies, in particular the spectral shape of the turbulence components. For example, the approach of \citet{Engelbrecht2017_TowardGreaterUnderstanding} that uses the NLGC theory can be shown to have a $\tau\Omega\propto \left(l_{sl}/l_{2D}\right)^{2/3}$ dependence: thus, the different values of this parameter, for example $l_{sl}/l_{2D}=10$ for \citet{Engelbrecht2017_TowardGreaterUnderstanding}, 2~in our study, and 1~in \citet{Ruffolo2012_RandomBallisticInterpretation}, may result in significant differences in the drift reduction factor. Further investigation on the sensitivity of the different perpendicular particle diffusion theories on different parameters is beyond the scope of this paper.

Our results demonstrate a sharp contrast to the predictions of the \citet{Bieber1997_PerpendicularDiffusionDrift} and \citet{Engelbrecht2017_TowardGreaterUnderstanding} results. At high energies, these predictions are similar to our results, however the energy dependence of our drift reduction factor is significantly weaker than that predicted by theoretical models based on field line or particle cross-field diffusion coefficients. It is quite possible that the issue is related to how the particle Larmor radius relates to perpendicular and parallel turbulence scales as discussed above. We will investigate this further in a forthcoming paper. An upcoming development is also to include the convective electric field $\mathbf{E}=-\mathbf{v}_{sw}\times \mathbf{B}$, where $\mathbf{v}_{sw}$ is the solar wind velocity, in our model. This will enable us to address drifts in longitude, for which the $\mathbf{E}\times\mathbf{B}$ drift is significant \citep[e.g.][]{Burns1968_DynamicsChargedParticle,Dalla2013_SolarEnergeticParticle}.

\section{Conclusions}

We have investigated the effect of magnetic field turbulence on the large-scale drifts present in the heliospheric magnetic field. As discussed in previous work, turbulence tends to reduce the amount of drift, and this has been characterised via a drift reduction coefficient $f_s$, which depends on particle properties such as energy and charge-to-mass ratio, and turbulence characteristics. Using test-particle full-orbit simulations, we have for the first time analysed the drift reduction due to turbulence in a heliospheric context, using our analytic heliospheric turbulence model \citep{Laitinen2023_AnalyticalModelTurbulence}. We found that
\begin{itemize}
    \item The drift reduction coefficient of protons is energy-dependent, with $f_s=0.2$ for 10~MeV proton in moderate turbulence with $dB^2/B^2=0.6$ at 1~au heliocentric distance. At higher energies, 100~MeV and 1000~MeV, the reduction is $f_s=0.7$ and 0.9, respectively.
    \item Stronger turbulence, with $dB^2/B^2=2$, gives rise to stronger drift reduction, with $f_s=0.2$ for 100~MeV protons, whereas for weaker turbulence, $dB^2/B^2=0.2$ the reduction is small, with $f_s=0.9$.
    \item Our values of the drift reduction coefficient are similar to those obtained by simulations in constant magnetic field by \citet{Minnie2007_SuppressionParticleDrifts} and \citet{Tautz2012_DriftCoefficientsCharged}. However, care should be taken with the comparison as the turbulence and background models differ significantly.
    \item According to our simulations, the drift reduction is significantly weaker than that proposed by theoretical models by \citet{Bieber1997_PerpendicularDiffusionDrift} and \citet{Engelbrecht2017_TowardGreaterUnderstanding}, particularly at lower energies. 
\end{itemize}

Thus, we find that while the turbulence does reduce the macroscopic drift in the IMF, the strong reduction predicted by theoretical approaches \citep{Bieber1997_PerpendicularDiffusionDrift, Engelbrecht2017_TowardGreaterUnderstanding, Vandenberg_2021_TurbulentReductionDrifts}, particularly at lower energies is not supported by our simulations. Thus, we expect that the effects of drifts on SEPs remain significant at large ion energies (e.g. $>100$~MeV protons), particularly for heavier elements which have larger Larmor radii.

\section*{Data Availability}

The simulation data used in this study are available as CSV files via Zenodo at \dataset[doi: 10.5281/zenodo.14284455]{https://doi.org/10.5281/zenodo.14284455}. Electronic versions of Tables 1 and 2 are also provided in CSV format.

\begin{acknowledgements}
TL and SD acknowledge support from the UK Science and Technology Facilities Council (STFC) through grants ST/V000934/1 and ST/Y002725/1. This work was performed using resources provided by the Cambridge Service for Data Driven Discovery (CSD3) operated by the University of Cambridge Research Computing Service (www.csd3.cam.ac.uk), provided by Dell EMC and Intel using Tier-2 funding from the Engineering and Physical Sciences Research Council (capital grant EP/P020259/1), and DiRAC funding from the Science and Technology Facilities Council (www.dirac.ac.uk).
\end{acknowledgements}

\bibliographystyle{aasjournal}
\bibliography{ms}

\begin{appendix}

\section{Turbulence parameters} \label{app:turbparams}

Our simulations use the same turbulence parameters as \citet{Laitinen2023_AnalyticalModelTurbulence}, aside from the turbulence amplitude which is varied, as given in column~7 of Table~\ref{tab:drifts}. The other parameters are briefly presented here for the reader's convenience.

The turbulence generation is based on a superposition of Fourier modes logarithmically equispaced in wave number $k$, using the approach by \citep{Giacalone1999_TransportCosmicRays}. We use a 2D-slab composite model with separate spectra for 2D ($k_\perp$) and slab ($k_\parallel$) components, with the power in the two components divided as 80\%:20\%. The power spectra consists of a large-scale $k^{p}$ component, with $p=0$, at wavenumbers below the breakpoint scales $l_{c\perp}$ and $l_{c\parallel}$, respectively (note that we use $l_c$ here instead of $\lambda_c$ which was used by \citep{Laitinen2023_AnalyticalModelTurbulence}, in order to avoid confusion with the parallel scattering mean free path $\lambda_\parallel$). At higher wavenumbers, we use Kolmogorov spectrum with power law index $8/3$ and $5/3$ for the 2D and slab components respectively. The breakpoint scales are defined as $l_{c\perp}=0.04\,(r/r_\odot)^{0.8}\;r_\odot$, where $r$ is the heliocentric distance and $r_\odot$ is the solar radius, and $l_{c\parallel}= 2 l_{c\perp}$. Finally, the total amplitude of the turbulence, $\delta B $, varies with location as $\delta B^2\propto r^{-3.3}$, with $\delta B^2/B^2=0.03$ at $r_\odot$, where $B$ is the background magnetic field given by Equation~(\ref{eq:parker}).

For further information about the parameters, their sources, and how they are used in the model, we refer the reader to the full description of the model in \citet{Laitinen2023_AnalyticalModelTurbulence}.

\section{Theoretical drift velocity at 1 au}\label{app:drift1au}

\begin{figure*}[ht]
    \centering
    \includegraphics{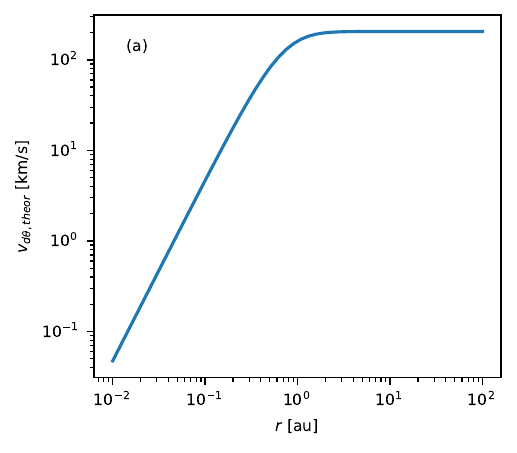}
    \includegraphics{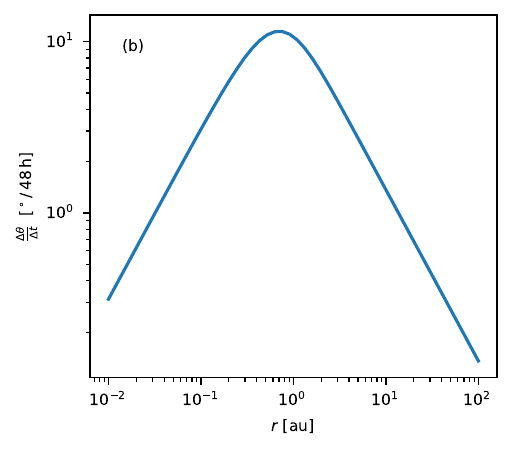}
    \caption{The theoretical drift velocity in heliolatitude, $v_{d\theta, theor}$, of 100 MeV protons in units of (a) km/s and (b) $^\circ/ 48\,\mathrm{h}$, as a function of heliocentric distance in the heliospheric equatorial plane, for the solar wind speed and Parker spiral magnetic field strength used in this study.}
    \label{fig:vdtheta_theor}
\end{figure*}

The pitch-angle dependent drift velocity in colatitude due to magnetic field gradient and curvature in Parker spiral is given by \citet{Dalla2013_SolarEnergeticParticle} as
\begin{equation}\label{eq:Dalladrift}
v_{d\theta, theor} = A\, \frac{\gamma m}{q} \left(\frac{1}{2}v_\perp^2+v_\parallel^2\right)f(r,\theta)
\end{equation}
where
\begin{equation}
  f(r, \theta)=\frac{a}{B_0 r_0^2}\frac{x^2(x^2+2)}{(x^2+1)^2}
\end{equation}
with $x=r/a(\theta)$ and $a(\theta)=v_{sw}/(\Omega\sin\theta)$, and $A$ is the sign of the Parker spiral magnetic field, as in Equation~(\ref{eq:parker}). Note that \citet{Dalla2013_SolarEnergeticParticle} show the drift rate in latitude rather than colatitude, hence their sign of $v_{d\theta}$ is opposite to ours. Our model is also unipolar unlike the \citet{Dalla2013_SolarEnergeticParticle} who used bipolar magnetic field, hence we do not incorporate the change of the sign of the drift velocity at the equator, $\mathrm{sgn}\left(\frac{\pi}{2}-\theta\right)$ in Equation~(\ref{eq:Dalladrift}).

To evaluate the representative value of the drift velocity, we average Equation~(\ref{eq:Dalladrift}) over the particle velocity distribution, assuming an isotropic pitch angle distribution. The isotropy assumption is justified by the fact that for our simulation cases, the scattering time scale at 1~au has values between $\sim 500$ and 2000 seconds, which are short compared to the timescales relevant to the determination of the drift velocity. Following the pitch-angle averaging, the terms in the parentheses in Equation~(\ref{eq:Dalladrift}) become 
\begin{equation}\nonumber
\overline{\frac{1}{2}v_\perp^2+v_\parallel^2}=\frac{2}{3} v^2.
\end{equation}
In our simulations, we characterise the Parker spiral with solar wind speed  $v_{sw}=400 \mathrm{\,km\,s^{-1}}$. Further, our region of interest is at the solar equator, with $\theta=90^\circ$, for which the factor
\begin{equation}\nonumber
f(r, \theta) = \frac{a_{eq}}{B_0}\frac{r^2}{r_0^2}\frac{r^2+2 a_{eq}^2}{(r^2+a_{eq}^2)^2}
\end{equation}
where $a_{eq}=a(\theta=90^\circ)\approx 0.93$~au, and the drift velocity 
\begin{equation} \label{eq:theor_drift_vel}
v_{d\theta, theor} = \frac{2}{3} A\, a\,r_{L0}\,v\frac{r^2}{r_0^2}\frac{r^2+2 a_{eq}^2}{(r^2+a_{eq}^2)^2}.
\end{equation}
where $r_{L0}$ is the particle Larmor radius at magnetic field of $B_0$. This can also be written in terms of the Larmor radius of the particle at 1~au as
\begin{equation}
v_{d\theta, theor} = \frac{2}{3} A\,r_{L}\,v\frac{r^2+2 a_{eq}^2}{(r^2+a_{eq}^2)^{3/2}},
\end{equation}
Using this $v_{d\theta, theor}$, the deviation in colatitude in the time interval $\Delta t$, at distance $r$, would be expected to be 
\begin{equation} \label{eq:theta_deviation}
    \Delta\theta = \frac{v_{d\theta} \Delta t}{r}.
\end{equation}
We demonstrate this in Figure~\ref{fig:vdtheta_theor}, where panel (a) shows the theoretical drift velocity in km/s, and panel (b) in angular units, for the parameters shown in this study.

It is important to note that in our simulations the particle's location will have deviated from $r=1$~au between its first and last crossing of the 1-au sphere. As discussed in \citet{Dalla2013_SolarEnergeticParticle} and shown in Figure~\ref{fig:vdtheta_theor} (a), the theoretical drift velocity due to curvature and gradient tends towards a constant value at large distances, $r\gg a$. Thus, as a consequence $r$ in the denominator of Equation~(\ref{eq:theta_deviation}), the deviation in colatitude, for a given $\Delta t$, decreases at small and large heliocentric distances, as shown in Figure~\ref{fig:vdtheta_theor} (b). Therefore, a particle that propagates at a wide range of heliocentric distances between its first and last crossing of the 1~au sphere will have drifted with smaller (angular) drift velocity on average than a particle that would have remained at 1~au.

For this reason, the drift velocity defined in Equation~(\ref{eq:vdtheta_definition}) can be expected to be smaller than $v_{d\theta, theor}$ for a particle that has propagated to small or large heliospheric distances before returning to 1~au.

\end{appendix}
 
\end{document}